\documentclass{article}
\usepackage{spconf,amsmath,amssymb,graphicx}
\usepackage[noadjust]{cite}

\title{POISONED FEEDBACK: THE IMPACT OF MALICIOUS USERS IN CLOSED-LOOP MULTIUSER MIMO SYSTEMS}
\name{Amitav Mukherjee and A. Lee Swindlehurst
\thanks{This work was supported by the U.S. Army Research Office under the Multi-University Research
Initiative (MURI) grant W911NF-07-1-0318.}}
\address{Dept.~of Electrical Engineering \& Computer Science\\
Henry Samueli School of Engineering\\
University of California Irvine\\
Irvine, CA 92697\\
\tt a.mukherjee@uci.edu
}
\begin{document}
%
\maketitle
\begin{abstract}
Accurate channel state information (CSI) at the transmitter is critical for maximizing spectral efficiency on the downlink of multi-antenna networks. In this work we analyze a novel form of physical layer attacks on such closed-loop wireless networks. Specifically, this paper considers the impact of deliberately inaccurate feedback by malicious users in a multiuser multicast system. Numerical results demonstrate the significant degradation in performance of closed-loop transmission schemes due to intentional feedback of false CSI by adversarial users.
\end{abstract}
\section{INTRODUCTION}
\label{sec:intro}
Effective interference management and spatial multiplexing of data in multiuser wireless systems is greatly dependent upon the accuracy of channel state information (CSI) at the transmitter. The use of feedback from receivers in multiuser wireless networks has now become a well-established technique to provide CSI at the transmitter \cite{Love08}. A number of analyses of imperfect feedback scenarios motivated by practical considerations are available, such as partial CSI feedback \cite{Madhow01}, limited-rate feedback \cite{Jindal06}, noisy feedback \cite{Milstein05}, and delayed feedback \cite{WCNC07}.

However, the problem considered in this paper is significantly different. Since the performance advantage of closed-loop transmission schemes over their open-loop counterparts is completely determined by the quality of the CSI, this opens the door to deliberate misreporting of CSI by malicious users as a novel form of a physical layer attack. Jamming and eavesdropping are the traditional categories of physical layer attacks in the literature, and have been widely studied for multi-antenna systems \cite{Mukherjee09}. To the author's best knowledge this is the first work to investigate physical layer attacks on MIMO systems based on malicious feedback of CSI.

 In particular, we examine malicious or \emph{poisoned} feedback attacks on the downlink of a multi-antenna network that is multicasting a common message to multiple receivers. The message being transmitted has no intrinsic value for the attacker; the malicious user is only interested in compromising the Quality-of-Service (QoS) provided to the legitimate receivers. In network security parlance, malicious behavior by authenticated users from within the network are referred to as `Byzantine attacks', and have usually been studied at the network and transport layers \cite{byzantine}.

 The remainder of this paper is organized as follows. The multicast network model and the adversarial user's capabilities are described in Sec.~\ref{sec:model}. The various forms of malicious feedback based on the corresponding objectives of the transmitter are listed in Sec.~\ref{sec:poison}. Numerical results that depict the impact of poisoned feedback are shown in Sec.~\ref{sec:simul}, and conclusions drawn in Sec.~\ref{sec:concl}.

\section{MATHEMATICAL MODEL}\label{sec:model}
The network under consideration is comprised of a $N_t$-antenna
transmitter multicasting to $\tilde K$ legitimate receivers and a single malicious user, all equipped with a single
antenna\footnote{It is straightforward to extend the principle of poisoned feedback to the case where each receiver is also equipped with an antenna array, for which multicasting strategies have been proposed in \cite{Boche08,Tomecki09}.} each, such that $\tilde K+1=K$ is the total number of receiving nodes.

In the general multicast scenario, a common scalar information symbol $z$ of unit power
is transmitted to all $K$ receivers. This necessitates the use of a common $N_t\times 1$
transmit beamformer $\mathbf{u}$ with with power constraint $||\mathbf{u}||_2^2 \leq P$. Compared to the broadcast scenario of independent information per receiver, the multicast beamforming problem was shown to be NP-hard \cite{Sidiropoulos06}. This led to the development of a number of approximate solutions based on techniques such as semidefinite programming, for instance \cite{Sidiropoulos06}-\cite{Boche08}.

 The $N_t\times 1$ transmitted signal is
\begin{equation}
{\mathbf{x}} = \mathbf{u}z.
\end{equation}
The received signals in a flat fading scenario are
\begin{equation}
{\mathbf{y}}_k  = {\mathbf{h}}_k {\mathbf{\mathbf{u}}}z +{{n}}_k, \quad k = 1, \ldots ,K,
\end{equation}
where ${\mathbf{h}}_k$ is the $1 \times N_t$ channel state vector for user $k$, and ${\mathbf{n}}_k$ is additive white Gaussian noise with variance $\sigma _k^2$. Due to the absence of inter-user interference, the signal-to-noise ratio (SNR) is the primary figure of merit:
\begin{equation}
\operatorname{SNR} _k  = \frac{{\mathbf{h}}_k {\mathbf{uu}}^H{\mathbf{h}}_k^H}
{{\sigma _k^2 }}.
\end{equation}

We focus on the following potential transmitter objectives in a multicast scenario:
\begin{enumerate}
\item Minimization of the transmit power subject to a minimum SNR threshold per receiver.
\item Maximization of the average received SNR for all receivers.
\item Maximization of the minimum user SNR (max-min) under the total power constraint $P$.
\item Maximization of the minimum information rate under the total power constraint $P$.
\end{enumerate}

 Objectives 3 and 4 are equivalent for the case of a single multicast group as in this work. To achieve any of the above system objectives, the transmitter requires global channel state information of all $K$ receivers ${\mathbf{H}} = \left[ {\begin{array}{*{20}c}
   {{\mathbf{h}}_1 } &  \ldots  & {{\mathbf{h}}_{\tilde K} } & {{\mathbf{h}}_a }  \\
 \end{array} } \right],$ where the subscript $a$ denotes the malicious adversary. On the other hand, the malicious user seeks to degrade the system performance objectives to the best of its ability by manipulating the CSI it feeds back.

 We assume that all $\tilde K$ legitimate receivers truthfully transmit their CSI to the transmitter over a error-free public feedback link. Moreover, this global CSI is also known to the malicious user via eavesdropping. The transmitter is assumed to be unaware of the presence of the malicious user and seeks to service all active receivers, i.e., user selection is not considered. The formulation of the resultant poisoned feedback ${{\mathbf{h}}_a }$ from the malicious user is described in the next section.

\section{POISONED FEEDBACK}\label{sec:poison}
\subsection{Transmit Power Minimization}
In this scenario, the transmitter seeks to minimize its transmit power required to satisfy a pre-determined minimum SNR target $\gamma$ for each receiver. On the other hand, the malicious user seeks to maximize the resource consumption at the transmitter. Towards this end, a crude attack would be to demand a very high QoS threshold relative to the legitimate receivers. However, such anomalous attacks are easy to identify, and at the very least would result in the malicious user being dropped from the set of scheduled receivers. Therefore, we consider a more subtle attacker, who seeks to feed back the worst possible channel state information so as to maximize the power consumption at the transmitter.

The malicious user has the following relaxed optimization problem:
\begin{equation}
\begin{gathered}
  \mathop {\max }\limits_{{\mathbf{h}}_e } \mathop {\min }\limits_{\mathbf{u}} \operatorname{trace} \left( {{\mathbf{uu}}^H } \right) \hfill \\
  s.t.\operatorname{trace} \left( {{\mathbf{uu}}^H {\mathbf{h}}_k {\mathbf{h}}_k^H } \right) \geqslant \gamma,{\text{ }}k = 1, \ldots ,K \hfill \\
 \hspace{0.6in} \operatorname{trace} \left( {{\mathbf{uu}}^H } \right) \leq P \hfill\\
  \|\mathbf{h}_a\|_2^2 \geq \beta,
\end{gathered}
\end{equation}
where an additional norm constraint has been placed on $\mathbf{h}_a$ by the attacker to avoid anomalous feedback values. Define ${\mathbf{D}} \triangleq {\mathbf{h}}_e {\mathbf{h}}_e^H$, ${\mathbf{U}} \triangleq {\mathbf{uu}}^H$, and ${\mathbf{G}}_k  \triangleq {\mathbf{h}}_k {\mathbf{h}}_k^H.$
Introducing an auxiliary variable $t$, we have the following SDP relaxation for the attacker:
\begin{equation}
\begin{gathered}
  \mathop {\min }\limits_{\mathbf{D}}  - t \hfill \\
  s.t.{\text{ }}\operatorname{trace} \left( {\mathbf{U}} \right) \geqslant t \hfill \\
  {\text{trace}}\left( {{\mathbf{UG}_k}} \right) \geqslant \gamma \hfill \\
  {\text{trace}}\left( {\mathbf{D}} \right) \geqslant \beta  \hfill \\
\end{gathered} \label{eq:SDP}
\end{equation}
Due to the relaxation of the rank-1 constraint on the transmit covariance, a randomization step is often required after the optimization in (\ref{eq:SDP}). This implies that the attacker may not be able to compute the same beamformer as the transmitter.
\subsection{Maximization of Average Received SNR}
Under this transmitter objective, the attacker adopts the following:
\[
\begin{gathered}
  \mathop {\min }\limits_{{\mathbf{h}}_a } \mathop {\max }\limits_{\mathbf{u}} \frac{{{\mathbf{uHH}}^H {\mathbf{u}}^H }}
{{\sigma _k^2 }} \hfill \\
  s.t.{\text{ }}\left\| {\mathbf{u}} \right\|_2^2  = P \hfill \\
\end{gathered}
\]
For the transmitter's maximization problem, a closed-form solution exists for the optimal beamformer $\mathbf{u}$, namely the principle eigenvector of ${\mathbf{HH}}^H$ \cite{Liu04}.

Intuitively, what the attacker should do here is to choose $\mathbf{h}_a$ to be very large and orthogonal to all of the other legitimate channel vectors.
The transmit beamformer would then approach $\mathbf{h}_a$, and all of the other users would see their allocated power approach zero.
\subsection{Maximization of Minimum SNR}
An alternative attack would be to minimize the maximum SINR enjoyed by any of the legitimate receivers.
\begin{equation}
\begin{gathered}
  \mathop {\min }\limits_{{\mathbf{h}}_e } \mathop {\max }\limits_{\mathbf{u}} \mathop {\min }\limits_{k} \operatorname{trace} \left( {{\mathbf{uu}}^H {\mathbf{h}}_k {\mathbf{h}}_k^H } \right) {\text{for }}k = 1, \ldots ,K,\hfill \\
  s.t. \operatorname{trace} \left( {{\mathbf{uu}}^H } \right) \leq P\\
  \|\mathbf{h}_a\|_2^2 \geq \beta.
\end{gathered}
\end{equation}

Broadly speaking, from the transmitter's perspective the optimal beamformer can be expressed as a linear combination of the user's channel state vectors:
\[
{\mathbf{u}}^H  = \sum\limits_{k = 1}^K {\alpha _k {\mathbf{h}}_k },
\]
where the complex coefficients $\alpha _k$ can be obtained using a sequential quadratic program \cite{Liu04}. 

However, instead of posing the above problem as another SQP or SDP which are known to be computationally intensive \cite{Utschick07}, we assume the attacker employs an iterative algorithm that alternatively optimizes $\mathbf{h}_a$ for a fixed $\mathbf{u}$, and vice versa. The inner optimization for the transmit beamformer can be carried out based on the iterative SNR-increasing update algorithm in \cite[Sec. VI]{Utschick07}. The attacker initializes the algorithm with an arbitrary channel vector and obtains the corresponding $\mathbf{u}$ for this initial global CSI matrix $\mathbf{H}$. After this step, the new candidate for $\mathbf{h}_a$ is obtained using a line search of appropriate step size in order to find the worst-case feedback in terms of the minimum SNR. These iterations continue until a pre-determined stopping criterion.

\section{NUMERICAL RESULTS}\label{sec:simul}
The following simulation results are compiled using 1000 Monte Carlo trials per point. The
channel vectors for all links are composed of independent Gaussian
random variables with zero mean and unit variance. The background
noise power is assumed to be the same for all $K$ receivers and the
eavesdropper: $\sigma_k^2=1$. All SNR and rate results shown here correspond to the $\tilde K$ legitimate receivers only, since the attacker has no value for the transmitted information as stated previously.

\begin{figure}[htbp]
\centering
\includegraphics[width=\linewidth]{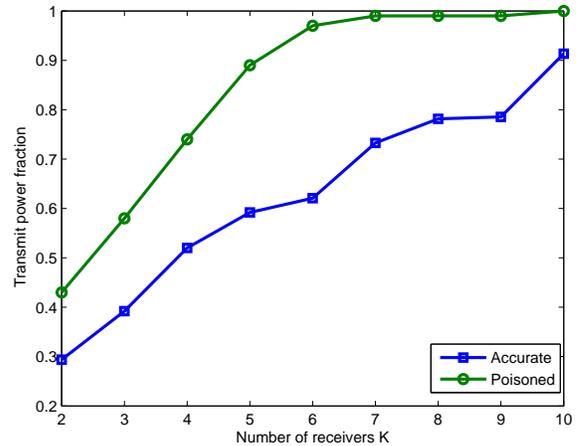}
\caption{Transmit power fraction versus number of receivers $K$, $P$=20dB, $N_t=5$ antennas.}
\label{fig_txpwr}
\end{figure}
Fig.~\ref{fig_txpwr} displays the contrast between the total transmit power required to meet a modest SNR target of $\gamma=5$dB per receiver when all receivers report their CSI accurately, and when a single malicious user is present. It is evident that the attacker is able to waste a significant portion of the transmitter's power.

\begin{figure}[htbp]
\centering
\includegraphics[width=\linewidth]{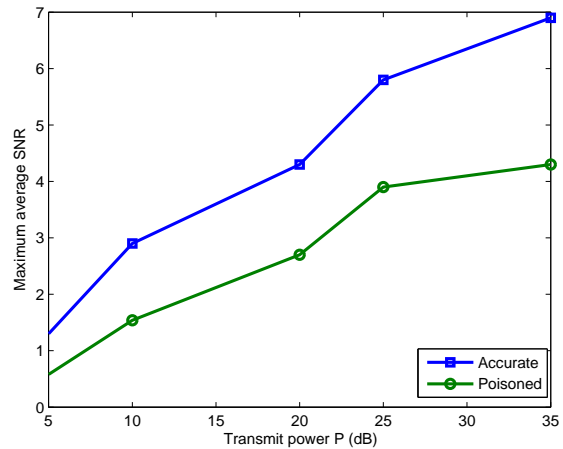}
\caption{Maximum average SNR versus transmit power $P$, $N_t=5$ antennas.}
\label{fig_maxavgSNR}
\end{figure}
Fig.~\ref{fig_maxavgSNR} exhibits the performance loss in terms of maximum average received SNR in dB of the legitimate users due to poisoned feedback, with $\tilde K=5$ receivers. The attacker is able to starve the other receivers of allocated power on the downlink, and reduces overall QoS levels by up to 3dB even for large transmit powers.

\begin{figure}[htbp]
\centering
\includegraphics[width=\linewidth]{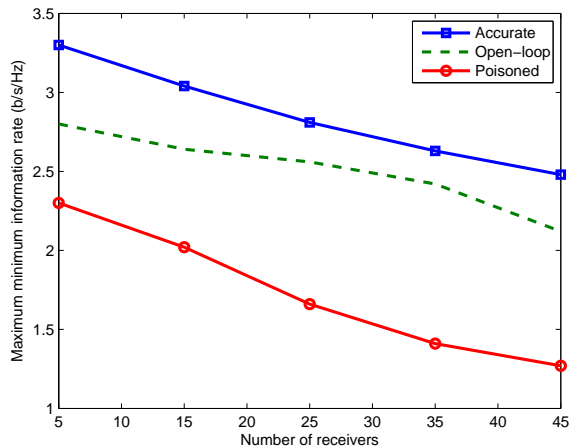}
\caption{Minimum information rate versus number of receivers $K$, $P$=20dB, $N_t=4$ antennas.}
\label{fig_rate}
\end{figure}
Fig.~\ref{fig_rate} shows the maximized minimum information rates for the closed-loop systems with completely accurate and poisoned feedback, and the open-loop multicast downlink with isotropic transmission \cite{Love08}, respectively. The maximized minimum information rate is defined as
\[
\mathop {\max }\limits_{\mathbf{D}} \mathop {\min }\limits_{1 \leqslant k \leqslant \tilde K} \log _2 \left( {1 + \operatorname{SNR} _k } \right).
\]
The interesting observation here is that the presence of just a single malicious user drives the system performance significantly below that achievable without any feedback whatsoever.
\section{CONCLUSION}\label{sec:concl}
This paper presented a preliminary investigation of the vulnerability of feedback-based downlink systems to malicious CSI reporting. It is observed that deliberate feedback of the worst possible CSI can lead to a closed-loop system performance that is considerably worse than that achieved by open-loop multicasting without CSI feedback. Therefore, smart detection and repudiation techniques to validate feedback of CSI at the physical layer are necessary as highlighted by the numerical results. Numerous avenues exist for future work, namely the closed-loop broadcast scenario with independent information for receivers.



\end{document}